\documentclass[preprint,12pt]{elsarticle}
\usepackage{amssymb}
\usepackage{amsmath}% http://ctan.org/pkg/amsmath
\usepackage{seqsplit}
\usepackage{lineno}
\usepackage{algorithm}% http://ctan.org/pkg/algorithms
\usepackage{algpseudocode}% http://ctan.org/pkg/algorithmicx
\usepackage{hyperref}
\usepackage{diagbox}
\newcounter{bla}

\journal{Computer Physics Communications}

\begin{document}

\begin{frontmatter}



%%\linenumbers

\title{A revision of the subtract-with-borrow random number generators}


\author{Alexei Sibidanov\corref{author}}

\cortext[author] {\textit{E-mail address:} sibid@uvic.ca}
\address{University of Victoria, Victoria, BC, Canada V8W 3P6}

\begin{abstract}
  The most popular and widely used subtract-with-borrow generator, also
  known as RANLUX, is reimplemented as a linear congruential generator
  using large integer arithmetic with the modulus size of 576 bits.
  Modern computers, as well as the specific structure of the
  modulus inferred from RANLUX, allow for the development of a fast modular
  multiplication -- the core of the procedure. This was previously
  believed to be slow and have too high cost in terms of computing
  resources. Our tests show a significant gain in generation speed
  which is comparable with other fast, high quality random number
  generators. An additional feature is the fast skipping of generator
  states leading to a seeding scheme which guarantees the uniqueness of
  random number sequences.
\end{abstract}


\end{frontmatter}

% Computer program descriptions should contain the following
% PROGRAM SUMMARY.

{\bf PROGRAM SUMMARY/NEW VERSION PROGRAM SUMMARY}
\begin{small}
\noindent
{\em Program Title: RANLUX++}                                 \\
{\em Licensing provisions: GPLv3}                             \\
{\em Programming language: C++, C, Assembler}                 \\
\end{small}
%%\linenumbers

\section{\label{intro}Introduction}
The well known Linear Congruential Generator (LCG) is a
recurrent sequence of numbers calculated as follows:
\begin{equation}
x_{i+1} = (a \cdot x_{i} + c) \bmod m,\label{eq:lcg}
\end{equation}
where $x_0$ is the initial state or {\it seed}, $a$ -- the multiplier,
$c$ -- the increment and $m$ -- the modulus. The particular choice of
the parameters $a$, $c$ and $m$ with period $q$ -- the minimal
number when $x_{q}=x_{0}$, can be found in the literature~\cite{Knuth:1997:ACP:270146}.
Commonly used LCGs are limited to $m \leq 2^{64}$, and have poor statistical
properties. Thus they are not used for Monte-Carlo physical simulations.

This situation can be mitigated when $m$ reaches several hundreds or
even thousand bits. The cost of the increased range of $m$ is to deal
with arbitrary precision integer arithmetic which was believed to be
prohibitively expensive for practical purposes. In the last two
decades there has been tremendous progress in modern central processor
units (CPU) especially for personal computers (PC) which can be
employed for long arithmetic.

We have explored the possibility to use the long arithmetic in LCG to
improve the quality of generated random numbers and found that,
despite a substantial increase in calculations, the time to generate a
single random number is not proportionally risen. In fact for some
parameters, the computational time decreased compared to ordinary LCGs
with machine word modulus size.

\section{\label{sb}Subtract-with-borrow generator}
At this point no specific constraints on $a$, $c$ and $m$ parameters
of LCG have been applied. As a good starting point we
choose the subtract-with-borrow generator first introduced
in~\cite{sbb} and the intimate connection with LCG has been shown as a
part of the period calculation. The algorithm has been extensively
studied in~\cite{LUSCHER1994100} to improve statistical quality of
generated numbers. Based on this study the generator
RANLUX~\cite{ranlux} was developed and now it is widely used in
physics simulations as well as in other fields where random numbers
with high statistical quality are required. However the current method
employed by RANLUX to achieve the high quality makes it one of the
slowest generators on the market.

The definition of the subtract-with-borrow generator is the following:
let $b$ some integer greater than 1 also called {\it the base} and
vector $Y = (y_1,\ldots,y_r,k)$ with the length $r+1$, where $0 \le
y_i <b$ and $k$ or {\it the carry} equals 0 or 1. Then define
a recursive transformation of the vector $Y_i$ with the rule:
\begin{equation}
  Y_{i+1} =
    \begin{cases}
      (y_2,\ldots,y_r,\Delta, 0), & \text{if}\ \Delta \ge 0\\
      (y_2,\ldots,y_r,\Delta + b, 1), & \text{otherwise}\\
    \end{cases}
    \label{eq:sbb}
\end{equation}
where $\Delta = y_{r-s+1} - y_{1} - k$ and $r$ and $s$ also called
{\it the lags}. As shown in the work~\cite{Tezuka}, this recursion is
equivalent to LCG with the modulus $m=b^r-b^s+1$, the multiplier
$a = m-(m-1)/b$ and $c=0$ with the relation:
\begin{equation}
  x_i = x(Y_i) \equiv \sum_{j=1}^{r}y_{j}b^{j-1}-\sum_{j=1}^{s}y_{r-s+j}b^{j-1} + k
\end{equation}

In the RANLUX generator the lags $r=24$ and $s=10$ with the base
$b=2^{24}$ are chosen among other suggested parameters in~\cite{sbb},
and thus the modulus $m = b^{24}- b^{10} + 1$ is a prime number and
the multiplier $a = m-(m-1)/b = b^{24} - b^{23} - b^{10} + b^{9} +
1$. With those parameters the period $q$ is equal to $(m-1)/48$.

Due to the selected base $b$ the natural choice to keep the
generator state is a vector of length 24 composed of 24-bit numbers.  This
implementation uses the properties of the modulus $m$ to avoid long
arithmetic calculations, and a single step equivalent to one modular
multiplication $(a\cdot x \bmod m)$ that requires only subtraction of two
24-bit numbers and carry propagation. In the original FORTRAN
implementation, 24-bit numbers were stored as floats to avoid at that
time, a high cost integer-to-float conversion.

\subsection{\label{rem}Remainder}
The simple structure of the modulus $m$ allows us to calculate the
remainder using only additions, subtractions and bit shifts. The
modulus $m$ and thus the generator state $x$ have size of $24\cdot 24 =
576$ bits and fits into 9 64-bit machine words.  The result of the
product $z = a\cdot x$ fits into 18 64-bit machine words which can be
represented as a 48 element array of 24-bit numbers: $z =
[z_{0},z_{1},\ldots,z_{46},z_{47}]$.  The number $r$ obtained by the
procedure shown in Algorithm~\ref{algorem} is congruent to $z \bmod m$
and $r < b^{24}$. Note the product $c \cdot m$ is also only bit
shifting due to the simple structure of $m$. The calculation of
$c$ is a sum of carry bits of each arithmetic operation.
\begin{algorithm}[t]
  \caption{Calculating remainder using only additions, subtractions
    and bit shifts for the modulus
    $m=b^{24}-b^{10}+1$. }\label{algorem}
  \begin{algorithmic}[1]
    \Procedure{Remainder}{$z$}\Comment{$0 \le z < b^{48}$}
    \State $t_0 \gets [z_{0} ,\ldots,z_{23}]$\Comment{$0\le t_0<b^{24}$}
    \State $t_1 \gets [z_{24},\ldots,z_{47}]$\Comment{$0\le t_1<b^{24}$}
    \State $t_2 \gets [z_{38},\ldots,z_{47}]$\Comment{$0\le t_2<b^{10}$}
    \State $t_3 \gets [z_{24},\ldots,z_{37}]$\Comment{$0\le t_3<b^{14}$}
    \State $r \gets t_0 - (t_1 + t_2) + (t_3 + t_2) \cdot b^{10}$
    \State $c \gets \lfloor r/b^{24} \rfloor$\Comment{$\text{floor}$ function rounds to $-\infty$}
    \State $r \gets r - c \cdot m$          \Comment{$0<r<b^{24}$}
    \State \textbf{return} $r$
    \EndProcedure
  \end{algorithmic}
\end{algorithm}

\subsection{Skipping}
Examining the result of a single step of Eq.~\ref{eq:lcg} one can note
that the main part of the number $x_i$ is preserved in its successor
$x_{i+1}$ which is just rotated by 24 bits. This strong correlation is the
reason of the poor statistical quality of the original
subtract-with-borrow generator~\cite{sbb}. The bright idea developed
in~\cite{LUSCHER1994100} is to apply the transformation~(\ref{eq:sbb})
many times to break correlations between nearby states before using
the state for actual physical simulation. The drawback of this method
is obvious -- all intermediate states have to be explicitly calculated
even if they are not needed. Despite the single step being simple with
small resource consumption, good statistical quality requires
several hundred steps thus in total, the skipping requires a lot of
time. This is a luxury to spend resources and not use the
results. Thus so-called luxury levels were introduced as aliases for
how many generatated numbers have to be wasted.

Using Eq.~\ref{eq:lcg} we can efficienty skip numbers since all $p$
recurrent steps collapse to a single multiplication:
\begin{equation}
  \underbrace{a \cdot (a \cdot (\ldots) \bmod m) \bmod m}_{p\ \text{times}}
  = (a^p \bmod m ) \cdot x \bmod m = A \cdot x \bmod m,
\end{equation}
where the factor $A \equiv (a^p \bmod m )$ is precomputed and thus the
cost to calculate the next state with or without skipping is the
same. Any state in the entire period $q=(m-1)/48\approx10^{171}$ can
be calculated in no more than $2\times\log_2(q) \approx 1140$ long
multiplications using fast exponentiation by squaring which takes
order of tens of $\mu$sec on modern CPUs.

In the Table~\ref{table:ap} the precomputed values of $A \equiv (a^p
\bmod m )$ where the values of $p$ is taken from~\cite{ranlux} are
shown for illustrative purposes. In the initial rows, long chains of 0
or 1 in binary representation are clearly visible and this can be
interepreted such that for each bit of the state $x_{i+p}$ only a few bits
of the state $x_i$ contributes. Even at the highest luxury level 4
there are still some patterns observable and a demading user maybe not
be completely satisfied. For such user the two last rows would be
more attractive especially since it is for free!  Such chaotic multipiers
mean that if any single bit of the state $x_i$ is changed in the next
step the altered state will be absolutely different from the unaltered
one.

With explicit long multiplication, there is no need to keep the
multiplier $A$ as a power of $a$, it can be adjusted to get the full
period, $m-1$. As an example the number $(a^{2048}+13 \bmod m)$ is a
primitive root modulo $m$ and with this multiplier all numbers in the
range $1\ldots m-1$ will appear in the sequence only once with any
initial $x_0$ from the same range.

\begin{table}
  %\tiny, \scriptsize, \footnotesize, \small, \normalsize (default), \large, \Large, \LARGE, \huge and \Huge. 
  \footnotesize
  \caption{\label{table:ap}The multiplier $A \equiv a^p \bmod m$ which corresponds
    to the RANLUX strategy to waste $p-24$ random numbers before next
    24 numbers will be delivered to a user as well as the associated
    luxury levels.}
  \begin{tabular}{p{0.05\linewidth}p{0.07\linewidth}p{0.83\linewidth}}
    luxury level & $p$ & $a^p \bmod m$ \\
    \texttt{0} & \texttt{  24} & \texttt{\seqsplit{fffffffffffffffffffffffefffffffffffffffffffffffffffffffffffffffffffffffffffffffffffe000000000000000000000001000000000000000000000000000000000000}} \\
    \texttt{1} & \texttt{  48} & \texttt{\seqsplit{000000000000000000000002ffffffffffffffffffffffff000000000000000000000000000000000001fffffffffffffffffffffffc000000000000000000000001000000000001}} \\
    \texttt{2} & \texttt{  97} & \texttt{\seqsplit{ffffff000000000008000000000009fffffffffffefffffffffff1000000000000000000000006ffffff000004fffffffffff6ffffffffffec000000000001000000000015000001}} \\
    \texttt{3} & \texttt{ 223} & \texttt{\seqsplit{00028b000000000bba00000000026cfffffffff8e4fffffffff96000000000027b0000000007d0fffffffffe25ffffffffeef0fffffffffa0a000000000942000000000ba6000000}} \\
    \texttt{4} & \texttt{ 389} & \texttt{\seqsplit{0df0600000002ee0020000000b9242ffffffdf6604ffffffe4ab160000000d92ab0000001e93f2fffffff593cfffffffb9c8a6ffffffe525740000002c38960000002ecac9000000}} \\
           $-$ & \texttt{1024} & \texttt{\seqsplit{e1754cefa19deea6f58651c8ac11b437ba841c49eca3003ff0ef508f058cfdab6105ca16980e6a3ab12a823219e1cd0007281433953609f1cc9c5ca19cf7f0c6d3899b14b7c5ee90}} \\    
           $-$ & \texttt{2048} & \texttt{\seqsplit{b48c187cf5b22097492edfcc0cc8e753ff74e54107684ed2256c3d3c662ea36c20b2ca60cb78c5096d8a15a13bee7cb0e64dcb31c48228ec4cec2c78af55c101ed7faa90747aaad9}} \\    
  \end{tabular}
\end{table}

The fast skipping also provides a seeding scheme which garantees
non-colliding sequences of random numbers -- what is needed is to just
skip a large enough sequence. This is suggested in~\cite{Tezuka} using the seed number as a power of the
multiplier:
\begin{equation}
  x_s = (a^n)^s \cdot x_0 \bmod m,
\end{equation}
where $s$ is the seed, $x_0$ is the initial state which is the same
for all seeds, $x_s$ is the starting state to deliver random numbers
for the seed $s$ and $n$ is the seed skipping factor to guarantee
non-colliding sequencies for any Monte Carlo simulation. Lets take
$n=2^{96}\approx10^{29}$ which is so big that for any modern computer
the required time to visit all states within the range exceeds the age
of the Universe, even if it can generate a new state on every CPU clock cycle. In
this case the maximum seed number is $s < q/n\approx
2^{474}\approx10^{143}$.

\section{\label{details}Implementation}
The core procedure for efficient implemention of LCG is fast long
multiplication. This provides the infrastructure to deliver random
numbers in an efficient as well as convinient form to users.
Programming languages such as C++, C or FORTRAN which are usually used
in Monte Carlo physics simulations, have no native support of the
long arithmetic despite all desktop grade CPUs providing hardware
instructions suitable for it. The support of long arithmetic is
provided by external libraries. We choose the GMP library~\cite{gmp}
for proof of concept. This library is highly optimized
for basic arithmetic operations with long numbers and supports a large
number of CPU architectures. It was found that GMP is a great tool but some
operations with long numbers still do not fit properly to the
interface provided and the conversion to a proper format and back adds
a significant overhead. This fact and for the sake of all inclusive
package without external library dependencies, the core functions are
written in the assembly language for the AMD64 CPU architecture. This
should fit to virtually all computers used for physics simulations. For
other architectures the long multiplication can be easily
reimplemented. The GMP version of the generator was used to validate
the resuts. The problem size is known in advance and this knowledge
has been exploited to write the fast and efficient procedures.

Since inception of AMD64 CPU architecture the instruction set has been
significantly expanded and new instructions profitable for long
arithmetic were introduced. Specificaly the instruction \texttt{mulx}
which streams the result of the multiplication to arbitrary registers
and then the instructions \texttt{adox} and \texttt{adcx} the addition
instructions which affect only corresponding overflow and carry flags
of the status register and are suitable to break the dependency chain
in long summations. To use advantages of the new instructions
where it is possible, three versions of 576 by 576 bit multiplication
were written and the best suitable version is selected and
executed in run-time by a CPU dispatcher. In all cases the method to
multiply numbers is a simple school-book method which requires 81
$64\times 64$ bit multiplications.

Considering the state $x$ as 576 bit entropy pool, to produce floating
point numbers, 24 or 52 bits are fetched from the pool for single or
double precision number correspondingly. The latter number has full
precision of 53 bits but from the 576 bits, only 10 53-bit numbers can be
produced, wasting the remaining 46 bits of entropy which are expensive. A
good compromise is to have random first 52 bits and waste only 4 bits
of entropy. Conversion of bit strings to the floating point
representation is done in batch into intermediate buffer from which
the numbers delivered to a user. This strategy significantly improves
overall performance.

\section{Benchmarks}
The benchmarks were conducted on three types of CPU -- Core2 (Intel(R)
Core(TM)2 Duo CPU E8400 @ 3.00GHz), Haswell (Intel(R) Core(TM)
i7-4790K CPU @ 4.00GHz) and Skylake(Intel(R) Core(TM) i5-6200U CPU @
2.30GHz).  Since data for all benchmarks fits to L1 cache and there is
not much dependence on main memory we present results of benchmarks in
CPU clocks.  It was observed that within CPU family benchmark results
are stable despite different CPUs can have largely different clock
frequencies. All measurements were done by the Linux {\tt perf} program which
provides convenient access to internal CPU performance counters.

The benchmark results of the specialized 9 by 9 64-bit limbs
multiplications together with the generic functions of GMP are shown in
Table~\ref{table:mult}. It is seen that modern CPUs provide significant
improvement in integer arithmetic especially in the straightforward
implementation of the long multiplication the author made. Apriory
knowledge of operand sizes gives additional performance boost compared
to the GMP functions which work with arbitrary size vectors.
\begin{table}
  \caption{\label{table:mult}CPU clocks needed to multiply two 576 bit
    numbers for generic functions from GMP (mul\_basecase\_*) and the
    author's $9\times9$ limbs multiplications for various CPU
    families.}
  \begin{tabular}{|c|c|c|c|}
    \hline
    \backslashbox{Function}{CPU type}
                            & Core2 & Haswell & Skylake \\ \hline
    mul\_basecase\_core2    & 370.9 & 218.5   & 197.1   \\
    mul\_basecase\_coreihwl & n.a.  & 205.6   & 203.7   \\
    mul\_basecase\_coreibwl & n.a.  & n.a.    & 163.0   \\
    mul9x9                  & 534.7 & 198.8   & 192.0   \\
    mul9x9mulx              & n.a.  & 162.1   & 154.5   \\
    mul9x9mulxadox          & n.a.  &  n.a.   & 119.4   \\
    \hline
  \end{tabular}
\end{table}

Algorithm~\ref{algorem} to calculate the remainder does not fit
properly into the GMP function set and no elegant and performance wise
ways were found, so only the author's implementations are tested. The
benchmarks of the modular multiplication are shown in
Table~\ref{table:mulmod} where the multiplication and the remainder
calculation are joined into a single function to keep intermediate
results in CPU registers. About 6.5-8.3 and 14-18 clocks is needed to
produce 24 and 52 bits of high quality entropy correspondingly for
single and double precision numbers.
\begin{table}
  \caption{\label{table:mulmod}CPU clocks needed for multiplication of two
    576 bit numbers modulo $2^{576}-2^{240}+1$.}
  \begin{tabular}{|c|c|c|}
    \hline
    \backslashbox{Function}{CPU type} & Haswell & Skylake \\ \hline
    mulmod9x9mulx           & 201.2 & 191.6  \\
    mulmod9x9mulxadox       &  n.a. & 155.4  \\
    \hline
  \end{tabular}
\end{table}

It is interesting to compare different implementations to perceive how
they can vary.  Fair comparison of different implementations of random
number generators requires special attention and microbenchmark
results on modern CPUs have to be cautiously interpreted due to
out-of-order execution of instructions, memory accesses in real
applications and so on. The small benchmark conducted is to sum $10^9$
random numbers uniformely distributed from 0 to 1. In this procedure
we ensured that the explicit \texttt{call} instruction is emmitted to
produce the random number in order to avoid inlining of a function body into
the summation loop. The results of testing various random number
generators are shown in Table~\ref{table:bench}.

The dummy function calculates nothing and simply immediately returns
0.5 demonstrating the overall overhead of the benchmark as well as its
uncertainty. The ``std'' prefix shows functions from the recent C++11
standard described in the $\langle$random$\rangle$
header~\cite{cpprandom} implemented in GNU C++ compiler. For the
subtract-with-borrow functions implemented in the C++11 standard the
double precision numbers, for the benchmarking purpose, have
explicitly only 48 bits of randomness to avoid the situation when 96
bits of entropy produced and only 53 bits of them are used. The
``gsl'' prefix shows functions from the GNU Scientific
Library~\cite{gsl}. TRandom1 is the ROOT implementation of
RANLUX~\cite{ROOT}. RANLUX is the original FORTRAN
implementation~\cite{ranlux}. The functions ranlxs and ranlxd are SIMD
implementations for single and double precision numbers
correspondingly of M.~L\"uscher, the author of the skipping
approach~\cite{LUSCHER1994100,ranlux33}. RANLUX++ is the suggested
implementation using the long modular multiplication.  The ``array''
comment shows another way to fetch produced random numbers -- first
fill a large array of numbers to avoid boundary checks for each number
and then sum numbers from the array.

The result of the benchmark shows that the suggested implementation of
LCGs with long modular multiplication is among the fastest random
numbers generators. The other RANLUX implementations with the high
statistical quality are an order of magnitude slower. In the current
benchmark the suggested generator is even faster than the simpliest
LCG in the C++ standard std::minstd\_rand. With already theoretically
proven statistical quality~\cite{LUSCHER1994100} also supported by
empirical tests~\cite{testu01} as well as the seeding scheme
guarantees non-colliding sequences this generator might be the best
option for large scale parallel physical simulations.
\begin{table}
  \caption{\label{table:bench}CPU clocks needed to sum $10^9$ random
    numbers at Haswell CPU normalized to a single number. All tests
    were compiled by gcc v6.2.1 with -O3 optimization flag.}
  \begin{tabular}{|l|c|c|}
    \hline
    \backslashbox{Generator}{Number type}
                          & double & float \\ \hline
    dummy                 &   9.1 &   9.1 \\ \hline
    std::minstd\_rand     &  35.1 &  20.2 \\ \hline
    std::mt19937\_64      &  36.0 &  37.0 \\ \hline
    std::ranlux24\_base   &  47.2 &  26.0 \\
    std::ranlux48\_base   &  24.4 &  26.1 \\
    std::ranlux24         & 387.0 & 197.7 \\
    std::ranlux48         & 640.5 & 638.4 \\
    gsl\_ranlxs0          &       &  84.2 \\
    gsl\_ranlxs1          &       & 125.9 \\
    gsl\_ranlxs2          &       & 215.5 \\
    gsl\_ranlxd1          & 213.8 &       \\
    gsl\_ranlxd2 ($p=397$) & 394.0 &       \\
    gsl\_ranlux ($p=223$) &       & 185.0 \\
    gsl\_ranlux389 ($p=389$)&       & 315.4 \\
    TRandom1 ($p=389$)    &       & 362.7 \\
    RANLUX ($p=389$)      &       & 378.3 \\
    ranlxs (array, SSE, $p=397$) &       &  50.4 \\
    ranlxd (array, SSE, $p=397$) &  95.0 &       \\ \hline
    RANLUX++              &  29.2 &  20.0 \\
    RANLUX++ (array)      &  24.7 &  15.7 \\
    \hline
  \end{tabular}
\end{table}

\section{Conclusion}
We revised the approach that the large moduli in linear congruental
generators are unpractical in MC physical simulations. We showed that
LCG can be effciently implemented in case of specially selected
modulus such as in the subtract-with-borrow generator. The famous
RANLUX random number generator was reimplemented using the long
modular multiplication.  The test results show an order of magnitude
improvement in the generation speed on modern CPUs. The generator has
a relatively small state vector of 72 bytes, it is fast and it has
already proven excellent statistical properties and thus can become
the best choice for large scale MC physical simulations. The source
code of the generator as well as some of the benchmark tests can be
obtained here~\cite{srcrepo}.

\appendix

\section{\label{effgen}An efficient implementation of the subtract-with-borrow generator}
The step of the subtract-with-borrow generator is very simple and the
current RANLUX algorithm can be implemented in a way to take
advantages of modern CPUs.  At the default luxury level $p=223$ about 
9 states are skipped so it is more efficient to skip an entire state, not
numbers one by one. This immediately gives a performance boost since
the array elements can be efficiently prefetched to CPU registers for
processing. The carry bit $k$ can be calculated exploiting the two's
compliment format of signed integer number. The carry $k$ in this
format is the most significant bit of the difference $\Delta$ which is
1 for $\Delta<0$ and 0 otherwise. Extracting the last 24 bits of the
difference $\Delta$ by bit masking makes the addition of $b=2^{24}$ in
Eq.~\ref{eq:sbb} redundant since it does not change those bits. The
optimized version of the step shown in Algorithm~\ref{algoopt} can be
easily implemented in the C language. The algorithm is also well
suited for a SIMD implementation similarly to what was done
in~\cite{ranlux33} running several generators in parallel according to
the width of SIMD registers. Several implementations of the algorithm
were written and tested. The test results are shown with $p=17\times
24=408$ (the closest to and greater than $p=389$ or the highest luxury
level 4 of the original RANLUX generator) in Table~\ref{table:opt} for
various types of optimization: ``scalar'' version is the
straightforward implementation of Algorithm~\ref{algoopt},
``scalar(asm)'' -- main loop is optimized in the assembler language to
access the hardware carry propagation, ``SSE2'' and ``AVX2'' are the
optimizations using vector instructions. The generation speed for the
latter case is so large that mainly the overhead of the benchmark
itself is seen. There is an order of magnitude speedup compared to the
results shown in Table~\ref{table:bench}.

The throughput of the skipping procedure itself without delivering
numbers to the user corresponds to what one would expect from
Algorithm~\ref{algoopt}: about 2~clock/(24 bit) for the scalar case,
0.5~clock/(24 bit) for SSE2 and 0.25~clock/(24 bit) for AVX2. The
assembler implementation of the scalar case was written to access the
hardware propagation of the carry bit, unfortunately the necessity to
clear 8 most significant bits of each number also affects the status
register with the carry bit. Thus this requires special treatment in the
code and only the throughput about 1.5~clock/(24 bit) was
achieved. These improvements can be easily applied in all current
implementations of the subtract-with-borrow generator.

\begin{algorithm}[t]
  \caption{Optimized version of the subtract-with-borrow step updating
    the element $x_i$ in $x=[x_0,\ldots,x_{23}]$ and the carry
    $k$.}\label{algoopt}
  \begin{algorithmic}[1]
    \Procedure{UpdateElement}{$i$, $j$, $k$}\Comment{$0\le i,j<24$, $0\le k\le1$}
    \State $d \gets x_{j} - x_{i} - k$ \Comment{32 bit signed integer in two's compliment format}
    \State $k \gets d\ \gg\ 31 $\Comment{logical right shift to extract the MSB}
    \State $x_i \gets d\ \&\ \texttt{0xffffff}$ \Comment{select and store the last 24 bits of $d$}
    \State \textbf{return} $k$\Comment{return carry}
    \EndProcedure
  \end{algorithmic}
\end{algorithm}

\begin{table}
  \caption{\label{table:opt}CPU clocks needed to sum $10^9$ single
    precision random numbers of the subtract-with-borrow algorithm
    with the skipping at Haswell CPU normalized to a single number. The
    skipping mode corresponds to the highest luxury level 4 of
    RANLUX. All tests were compiled by gcc v6.2.1 with -O3
    optimization flag.}
  \begin{tabular}{|c|c|c|c|}
    \hline
    scalar & scalar(asm) & SSE2 & AVX2 \\
    \hline
    40.3 & 28.5 & 12.0 & 7.95 \\
    \hline
  \end{tabular}
\end{table}

%% References


\bibliography{term}
\bibliographystyle{elsarticle-num}



 %% \begin{thebibliography}{00}
 %% \end{thebibliography}

\end{document}